# Validation of Rigorous Requirements Specifications and Document Automation with the ITLingo RSL Language

André Rodrigues[1] and Alberto Rodrigues da Silva[2]
[1]Instituto Superior Técnico, Universidade de Lisboa, Lisboa, Portugal
[2]INESC-ID, Instituto Superior Técnico, Universidade de Lisboa, Lisboa, Portugal
{andre.f.mota.rodrigues, alberto.silva}@tecnico.ulisboa.pt

**ABSTRACT**

Despite being an essential step in software development, writing requirements specifications is frequently performed in natural language, leading to issues like inconsistency, incompleteness, or ambiguity. The ITLingo initiative has introduced a requirements specification language named RSL to enhance the rigor and consistency of technical documentation. On the other hand, natural language processing (NLP) is a field that has been supporting the automatic analysis of requirements by helping to detect issues that may be difficult to see during a manual review. Once the requirements specifications are validated, it is important to automate the generation of documents for these specifications to reduce manual work, reduce errors, and to produce documentation in multiple formats that are more easily reusable or recognized by the different stakeholders. This paper reviews existing research and tools in the fields of requirements validation and document automation. We propose to extend RSL with validation of specifications based on customized checks, and on linguistic rules dynamically defined in the RSL itself. In addition, we also propose the automatic generation of documents from these specifications to JSON, TXT, or other file formats using template files. We use a fictitious business information system to support the explanation and to demonstrate how these validation checks can assist in writing better requirements specifications and then generate documents in multiple formats based on them. Finally, we evaluate the usability of the proposed validation and document automation approach through a user session.

**KEYWORDS**

Requirements engineering, Requirements validation, Requirements specification, Document automation, Natural language processing, ITLingo RSL

## 1 INTRODUCTION

Requirements specifications are essential when specifying system requirements. Unambiguous and consistent requirements are crucial for the success of IT projects as the later these errors are detected, the higher the cost associated with them [1]. Thus, it is important to clearly identify and validate requirements early in the project [2]. However, the requirements analysis and validation process tends to be expensive, subjective, and error-prone [2], [3]. So, it is crucial to standardize requirements and automate their validation to improve quality. It is recommended in this process to translate requirements written in a natural language into a controlled natural language (CNL) that facilitates validation while maintaining some level of understanding among stakeholders [4]–[6]. Several key techniques are used in requirements engineering, namely elicitation, specification, verification and validation, and management to ensure that requirements satisfy the needs of stakeholders [1].

In this verification and validation process, NLP techniques can help examine the natural language used in requirements and spot common errors like inconsistencies or ambiguities [7].

ITLingo is a research initiative that intends to improve the quality and efficiency of how engineers and domain experts produce and manage technical documentation in the IT domain [2] following a model-driven engineering (MDE) approach [8], which has been expanded from an Eclipse-based tool called ITLingo-Studio to a web version, the ITLingo-Cloud [9]. In the scope of ITLingo, some rigorous languages have been designed and discussed in previous studies [2], such as the RSL (Requirements Specification Language) which has been designed to support the production of system requirements specification in a more systematic, rigorous, and consistent way [4], [10]–[12].

Although RSL is a controlled natural language with a limited vocabulary and grammar, making the specifications easier to read and understand by computers [4], the existing validation is only performed at the grammatical level by checking the rules of the language concepts. Therefore, it is important to have more advanced and flexible options covering aspects not considered at the grammatical level and, simultaneously, be a learning tool that gives continuous feedback to the user writing the requirements specifications. Once these specifications have been validated, it is also crucial to convert them into technical documents that are easier for all stakeholders to understand.

This paper reviews some tools and research conducted in the requirements validation and document automation process. It discusses extra validations on those RSL specifications for the ITLingo-Cloud's web IDE by introducing customized checks in the language's validator and validations based on a fragment of the language itself named linguistic rules. Once validated, our approach enables the generation of documents from these RSL specifications in multiple formats, as illustrated in Figure 1.

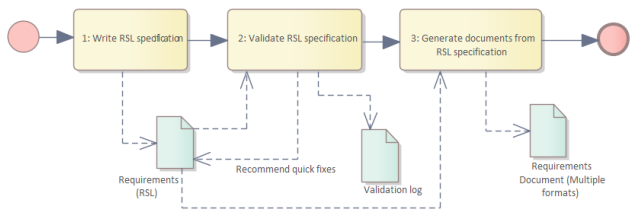

**Figure 1: Validation and generation of documents based on RSL specifications**

This paper is structured in 7 sections. Section 2 introduces the concepts inherent to the conducted research. Section 3 introduces the validation based on customized checks and linguistic rules approach. Section 4 describes our solution for document automation from specifications. Section 5 discusses these validations and document automation techniques with the support of a fictitious business information system and a usability assessment performed by users. Section 6 presents and discusses the existing research and tools for requirements validation and document automation. Section 7 summarizes the main contributions and presents future work.

## 2 BACKGROUND

This section reviews the key concepts relevant to our work regarding requirements validation and document automation, along with a brief introduction to the ITLingo RSL and the tools relevant to this research.

### 2.1 ITLingo and RSL

ITLingo RSL is a controlled natural language that helps the production of both requirements and test specifications systematically, rigorously, and consistently [4], [10]–[12]. RSL is a language that includes a rich set of constructs logically arranged into views according to multiple RE-specific concerns at business and system abstraction levels. These constructs are defined as linguistic patterns and grammatical rules enabling their users to speak correctly in a common language. They are represented textually by mandatory and optional fragments or text snippets [4].

RSL was initially developed using the Xtext framework [13], built on the Eclipse IDE that supports the definition of domain-specific languages (DSLs), and more recently using the Langium[1] framework built on Visual Studio Code, which is the web's equivalent of Xtext. These DSL frameworks contain techniques within the validation and error correction that allow embedding additional constraint checks and quick fixes in the grammar specification. In addition, both frameworks also include a large set of features for code generation and testing the implemented DSL.

**DataEntity e_Invoice** "Invoice": Document [
    attribute ID "Invoice ID": Integer [constraints (PrimaryKey)]
    attribute creationDate "Creation Date": Date [defaultValue "today"]
    attribute approvalDate "Approval Date": Date
    attribute paidDate "Payment Date": Date
    […] ]

**Actor a_Manager** "Manager": User [description "…"]

**UseCase uc_2_BrowseInvoicesToApprove**: EntitiesBrowse [
    primaryActor a_Manager
    dataEntity ec_Invoice
    actions aClose, aSearch, aFilter
    extensionPoints xp_ApproveInvoice]

**Spec 1: Simple RSL specification**

Spec. 1 illustrates a simple example of an RSL specification that defines three elements: (1) the data entity "Invoice" with several attributes; (2) the user actor "Manager" responsible for approving invoices; and (3) the use case "Browse invoices to approve" where the manager browses invoices and approves or rejects them.

### 2.2 Requirements Validation

An essential step in the software development process is the writing of requirements. Eliminating problems like ambiguity, inconsistency, or incompleteness is important, as these issues might result in misunderstandings, additional costs, or flaws in a system's functionality [1].

Requirements validation aims to align specifications with the system description, ensuring the specifications are comprehensive, consistent, compliant with relevant standards, free of inconsistencies and technical errors, and unambiguous [14].

Manual validation, due to its subjectivity and human dependence, can be expensive, time-consuming, and error-prone. Moreover, manual validation may not detect errors inherent to human language, particularly in highly intricate requirements, adding complexity to the validation process [1], [14]. As a result, it is essential to look for techniques to standardize the process of writing requirements, as performed in the ITLingo initiative with the RSL language, and to use automatic requirements analysis approaches to verify requirements more effectively.

---

[1] https://www.langium.org

Requirements specifications are considered completed when they cover a system's intended functionality. Requirements are consistent when they do not conflict with each other, which can make development difficult because unless any investigation is performed, one might not be able to tell which (if any) requirement is consistent.

Furthermore, requirements are unambiguous if every requirement has only one interpretation. Achieving unambiguity can be complex due to the inherent ambiguity of natural language, which can introduce unintended multiple meanings. To minimize ambiguity, formal or semi-formal specification languages are recommended as an alternative or complement to natural language.

For example, when using a controlled language like RSL, intended to minimize common errors, it is still possible to have errors such as (1) different names for the same concept, e.g., "Client" and "Customer"; (2) loops in relationships between elements, e.g., element "User" is an "Administrator" and "Administrator" is a "User". Therefore, these errors can affect the correctness and consistency of the requirements, resulting in problems in the System's interpretation and functionality.

## 2.3 NLP Technologies

Natural language processing (NLP) is a field of artificial intelligence that deals with computer-human interaction to achieve human-like processing for various tasks or applications such as speech recognition, natural language understanding, and natural language generation [15].

In requirements engineering, it is common to use NLP techniques to help identify and extract information from textual descriptions [7]. Among the different techniques, the ones relevant to this research are tokenization, part-of-speech (POS) tagging, and lemmatization.

Stanford CoreNLP [16] and Apache OpenNLP[2] are Java-based toolkits supporting several NLP tasks, such as tokenization, sentence segmentation, part-of-speech tagging, and named entity extraction. Natural Language Toolkit (NLTK) [17], a Python toolkit, provides easy-to-use interfaces to corpora and lexical resources such as WordNet[3] and text processing libraries. Spacy[4], another Python library, is designed for production use and is useful for building information extraction, natural language understanding systems and pre-processing text for deep learning.

Compromise[5] and WinkNLP[6] are JavaScript libraries designed to help the development of NLP applications. While Compromise provides functions for common NLP tasks in many languages to help manipulate text in a handy and uncomplicated way, WinkNLP features an extensive NLP pipeline and aims for the right balance between performance and accuracy in Node.js.

For the NLP tasks performed in this research, which included POS tagging in multiple languages, we used the Compromise library to support our research, given its easy integration with the Typescript-based Langium framework and its multilingual support.

## 2.3 Document Automation

During the requirements specification process, the task of writing technical documentation with these requirements can be challenging and time-consuming. Therefore, it is important to automate this process in addition to the validation process to accelerate and improve technical documentation [2].

As part of the ITLingo initiative, several M2T methods and techniques for transforming specifications into other representations and formats have already been discussed and developed for the Eclipse-based tool, the ITLingo-Studio [2].

The need to move these M2T transformations to a cloud IDE environment also arose with the expansion of the ITLingo initiative to the ITLingo-Cloud environment. As such, in the context of web-based applications, we discuss three template engines for automating document generation with the ITLingo RSL language, the handlebars [7], the EJS [8], and the docxtemplater[9].

Handlebars is an open-source JavaScript template engine that enables the creation of templates that closely resemble regular text with embedded expressions that get replaced by data from an input object.

Embedded JavaScript (EJS) is an open-source JavaScript template engine using only plain JavaScript code, ideal for those who want to build their template logic using it.

Docxtemplater is a library offering extensive customization designed to generate documents in formats such as MS Word (DOCX) and MS PowerPoint (PPTX), based on template documents in the same format using template tags.

Creating a template document user-friendly with minimal technical knowledge is crucial. However, EJS can be challenging when writing templates, as it requires expertise of the JavaScript language. Although Handlebars is a popular and free template engine, the docxtemplater offers extensive customization and already provides access to the MS Word module in the free version. Given that the technical documentation is typically authored in documents like MS Word and given its support for the MS Office file formats, we chose docxtemplater for our research, allowing an easier integration for anyone familiar with MS Word documents.

---

[2] https://opennlp.apache.org/
[3] https://wordnet.princeton.edu
[4] https://spacy.io/
[5] https://observablehq.com/@spencermountain/nlp-compromise
[6] https://winkjs.org/wink-nlp
[7] https://handlebarsjs.com
[8] https://ejs.co
[9] https://docxtemplater.com

# 3 REQUIREMENTS VALIDATION OF RSL SPECIFICATIONS

Some validations of RSL specifications are already supported by the Langium framework, which already provides automatic grammar validation. However, these validations are limited, and therefore, we have expanded them by implementing custom validations for RSL that are not possible by only checking the grammar. In addition, we have developed a flexible validation approach using elements of the language itself, named linguistic rules, to make it possible to validate fragments of RSL elements such as the ID, name, or description, i.e., whenever it is required to insert a text. This latter validation type is extremely important because the element's fragments are string-based and as such it is crucial to define rules to standardize the writing style, ensuring consistency and clarity in specifications.

These validations are performed on-the-fly, (i.e., while the user is typing) to help detect requirements that don't follow the required rules and suggest the respective correction. These corrections or quick fixes can be made in two ways: suggesting the automatic fix to the user or suggesting how the user can fix the error.

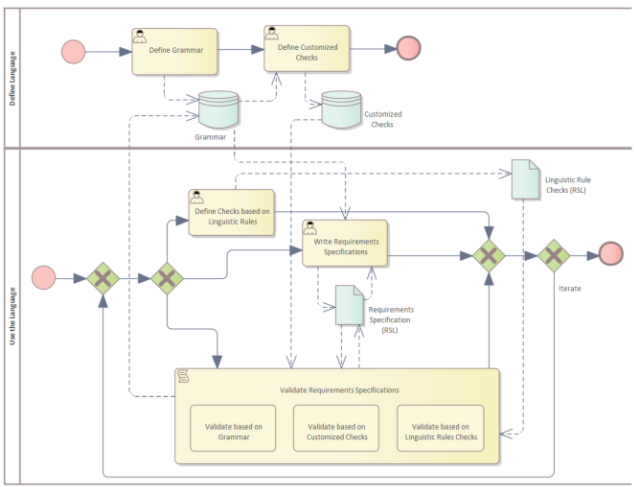

**Figure 2: Overview of the validation of RSL specifications based on grammar, customized checks, and linguistic rules**

In this validation process, as illustrated in Figure 2, we propose and discuss two types of validations complementing existing grammar validations: (1) customized validations defined by extending specific methods when the language is created using the framework language validator and quick fix provider; and (2) validations based on linguistic rules defined when using the language, checking for example if the ID, name or description fragments of an element follow some predefined rules.

## 3.1 Customized Validations

As introduced in the previous sections, the Langium framework supports grammar-level validations of RSL specifications. These validations ensure that language elements are valid according to a specified syntax and generate default error messages, as shown in Figure 3.

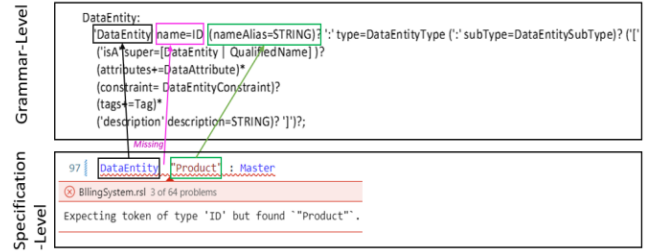

**Figure 3: Example of grammar validation**

Although an element can be syntactically correct but not valid, i.e., meaning it does not respect the semantics of the language; e.g., the same name for different elements.

The customized validation checks allow the implementation of a set of checks for elements that are not possible at grammar creation time. They also provide at the same time the generation of customized error or warning messages to the user.

This way, it is possible to perform additional validations to requirements specifications such as: (1) guarantee that element IDs are unique; (2) guarantee consistency in the use of terms and (3) prevent loops in hierarchical relationships. Other validations could be considered as well, such as checking whether an element's subtype is consistent with its supertype (each type only has specific subtypes) or checking whether the type of an element is consistent with its fragments (e.g., an active flow specification with conditions but is not of the "sequence conditional" type).

### 3.1.1 Guarantee that Element IDs are unique

The first validation checks that all element IDs in the requirements specifications are unique across the specification document, regardless of the element type.

For instance, as shown in Figure 4, when two actors and one data entity are defined with the same ID, "user", the System detects these defects and notifies the user by suggesting the creation of unique IDs.

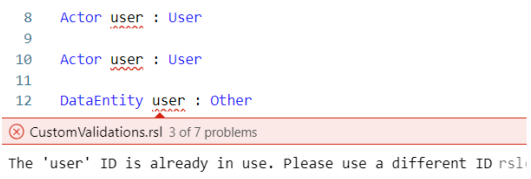

**Figure 4: Validation of duplicated element IDs**

### 3.1.2 Guarantee consistency in the use of terms

The second validation is performed when defining a glossary of terms (domain keywords) that shall be used consistently across RSL specifications to ensure that all element specifications use the same terminology.

As an example of the validation of synonyms defined in a glossary of terms, the RSL sentence Term t_Customer "Customer": Noun [synonyms "Client"] indicates that the word "Client" is synonymous of "Customer", and consequently, if any element description is found with the word "Client", suggests its replacement by the main word "Customer", as illustrated in Figure 5.

```
24    Term t_Customer "Customer" : Noun [synonyms "Client"]
25
26    Actor a_Customer "User that is a client" : User
⚠ CustomValidations.rsl 1 of 1 problem
Warning: Replace the word 'Client' by the main word 'Customer'
```

**Figure 5: Validation of terms defined in the glossary**

### 3.1.3 Prevent loops in hierarchical relationships

The third validation is applied when specifying an element's "is-a" or "part-of" relationship. When defining that an element has a relationship with another element, it is necessary to ensure no loop exists between them.

For example, when defining two actor elements, "Customer" and "Customer VIP", when it is indicated that a "Customer VIP" is a "Customer" actor and that a "Customer" is a "Customer VIP" actor, the System shall detect such loops in hierarchical relationships and generate an error to alert the user of this issue, as shown in Figure 6.

```
30    Actor a_Customer "Customer" : User [ isA a_CustomerVIP ]
31
32    Actor a_CustomerVIP "Customer VIP" : User [ isA a_Customer ]
⊗ CustomValidations.rsl 5 of 9 problems
Error: Cycle in hierarchy of Actor 'a_CustomerVIP' rsl(org.itlingo.
```

**Figure 6: Validation of hierarchical relationships**

## 3.2 Validation based on Linguistic Rules

Linguistic rules are syntactic rules that allow checking if some RSL element's fragments (e.g., ID, name, or description fragments) are properly written according to some linguistic pattern. This validation approach is supported by the Compromise library, which helps to check the correctness of these fragments and alert the user of any inconsistency or incompleteness.

The validation mechanism may also suggest some corresponding corrections, such as the creation of an element with a given name or description as illustrated in Figure 7, where the validation detects that the word "invoice" does not match any RSL element name, suggesting as a correction the creation of a data entity with the name "invoice". The main benefit of using linguistic rules compared to the previous types of validation is their ease and ad-hoc creation, as well as the fact that they can later be shared and replaced without having to change the validation mechanism.

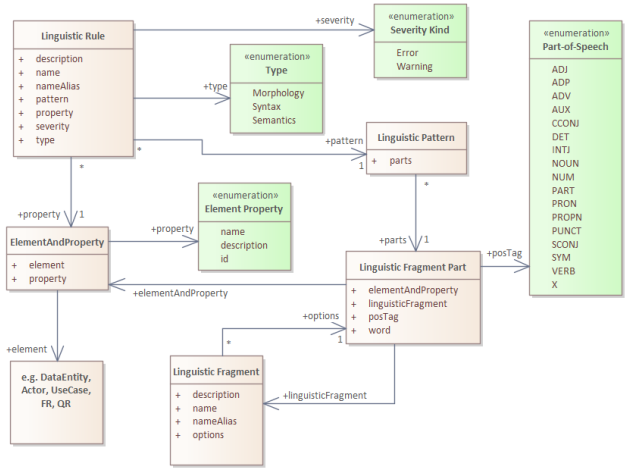

**Figure 7: Validation and corresponding quick fix suggestion based on a linguistic rule**

### 3.2.1 Linguistic Rules Specifications

The specification of a linguistic rule defines what validation will be performed and what linguistic pattern is suggested to the user in case the rule is not met. Figure 8 shows the metamodel of a linguistic rule as defined in the RSL grammar.

**Figure 8: RSL linguistic rule metamodel**

For example, the linguistic pattern *Verb + (DataEntity.name)* used for user case names (Spec 2) requires: (1) the sentence to begin with a verb; (2) followed by a data entity's name already present in the system.

**LinguisticRule LR_1** "Use Case name": Syntax [
  property UseCase.name
  pattern Verb + (DateEntity.name)
  severity Error
  description "Use case names must contain one verb and one data entity name"
]

**Spec 2: Example of a linguistic rule**

One significant advantage of the validation based on linguistic rules is that it facilitates validation in multiple languages, by merely specifying the natural language used in the specifications document. Therefore, the linguistic rules are also supported by an RSL element named "LinguisticLanguage". This element refers to the language used to write the ID, name, and descriptions of the RSL elements in the specifications document and therefore designates the language to which the linguistic rules are subsequently applied, which includes the following languages: English, Spanish, German, French, Italian, Portuguese, and Japanese.

Figure 9 shows an example of the validation of a use case name, utilizing Portuguese as the natural language. The validation identifies that the word "Criar" is a verb and corresponds to the first part of the pattern "(Verb) + (DataEntity.name)". However, there is no existing data entity named "Fatura", thereby suggesting its creation.

**Figure 9: Validation of a use case name based on the Portuguese language**

### 3.2.2 Linguistic Rules Validation

The validation of specifications based on linguistic rules is supported by the Compromise library, using key NLP techniques including sentence splitting, tokenization, part-of-speech tagging, and lemmatization that we use to deconstruct the sentence, collect the POS, and the lemmas of each word to be used in our validation process.

This validation process includes two stages: (1) a syntactic analysis of the sentences performed by the Compromise library; and (2) an analysis of the linguistic rules that have been previously defined. Figure 10 overviews this validation approach.

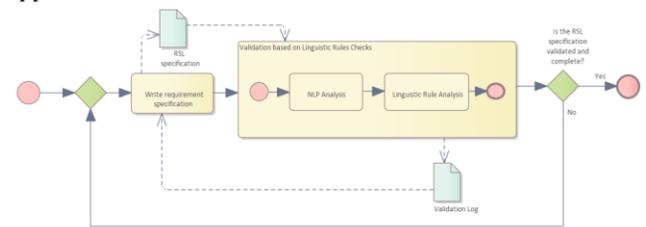

**Figure 10: Overview of the validation based on linguistic rules checks**

During the first stage, the respective element's ID, name, or description is analyzed by the Compromise library considering the natural language defined in the document. It then returns a token (word) with linguistic details, including the part of speech in the Universal Dependencies (UD) format and the lemma for each word.

In the second stage, our approach involves verifying for each RSL element the linguistic rule associated with the ID, name, or description of each respective element. In case the validation finds an RSL element that does not comply with the configured linguistic rule, the user is alerted about the conceptual schema of the rule that was not complied with using the configured validation severity level and gives suggestions to correct the text.

### 3.3 Specifications Library

As previously mentioned, one of the key benefits of using the linguistic rules is that once a set of linguistic rules has been defined, they can be shared and modified without the need to make changes to the validation mechanism.

RSL supports the reuse of specifications by allowing to include one or more elements from other packages and systems using the "Import", "Include" and "IncludeAll" elements.

Once these "include" specifications are defined, the elements they reference become part of the document being edited. For this reason, an informative quick fix suggests replacing these "include" specifications with the referenced specifications as illustrated in Figure 11.

**Figure 11: Quick fix generated for "Include" specifications**

## 4 DOCUMENT AUTOMATION BASED ON RSL SPECIFICATIONS

This section presents the techniques developed in the context of the ITLingo-Cloud related to the task of document automation.

As previously discussed, the use of a DSL like the RSL language arises from the need to reduce errors and improve efficiency in the creation of requirements. Once the requirements have been specified in a more controlled natural language and properly validated, it becomes important to generate technical documentation in more commonly used formats; for instance, with a more natural and simpler style, i.e., more easily understood by business stakeholders.

Therefore, we research three techniques of generating documents from RSL specifications: (1) the generation of JavaScript Object Notation (JSON) files; (2) the generation of plain text documents with a predefined structure for RSL elements; and (3) the generation of documents using template files, allowing to produce more complex and versatile documents in formats such as DOCX, or PPTX.

These generations were created with the assistance of the Langium framework in the VSCode editor. Langium includes an API to support the generation of text from DSL elements, facilitating the development of our document automation techniques. In addition, given that the RSL language IDE is generated in VSCode, we leveraged the VSCode's editor experience and added new context menu options and new interactive dialogs in the editor to enhance the user experience in the context of document automation.

As illustrated in Figure 12, for each option, the specifications shall be valid, and the user shall be alerted if the specifications have errors (either grammatical or errors identified by the validations introduced in this paper). In case the specifications file is not valid, the user is notified to rectify them before proceeding to generate a new file.

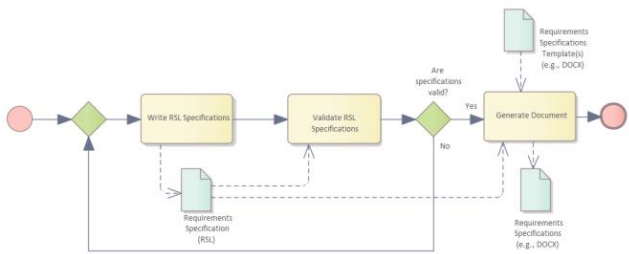

**Figure 12: Document automation approach**

The first option enables generating a JSON file from a specification. This generation offers several advantages, making it possible to integrate the RSL specifications with other formats and systems, due to its simplicity and ability to structure data easily and clearly.

The second option is the generation of plain text files from an RSL specification file. Every type of element (e.g., stakeholders, actors, use cases) is mapped to a more structured but still simple format, allowing to organize the data and have a less verbose text (without the rules of the language grammar).

The third option involves generating documents using template files with the help of the docxtemplater template engine. These template files can be defined in files with the extensions "docx", "pptx", "txt", "docm", "dotx", "dotm", "pptm". In our approach, we support the default tags of the docxtemplater engine, such as {actor} to insert the object data "actor".

To make this mechanism even more complete, we also extended the available tags with additional functionalities. Therefore, we added support for using angular expressions[10] in the template tags so that it would be possible to introduce more complex tags, opening an infinite number of possibilities.

## 5 EVALUATION

In this section, we evaluate the solution in the context of the validations introduced in the RSL language and the automation of documents from specifications with the help of the Billing System experiment. In addition, we present the results of a usability session with users that evaluated the usability of the introduced validation and document automation techniques.

### 5.1 The Billing System experiment

To better describe the concepts introduced previously and support the discussion, this research uses the requirements specification of a fictitious business information system, previously introduced by Silva [4], where specific errors were added to simulate issues that could be validated and addressed using the quick fix suggestion for the validation checks discussed in Section 3, and then generate documents from these validated specifications.

#### 5.1.1 Requirements Validation

To support the discussion of the validations presented in sections 3.1 and 3.2, we added customized-based issues and linguistic rules-based issues in the Billing System specification, namely: (1) duplicate element IDs; (2) alternative terms defined in the glossary; (3) loops in hierarchical relationships; (4) actor's name not complying with the linguistic rule; and (5) functional requirement description not complying with the linguistic rule.

The validations introduced with the customized checks allow the definition of validations that make the elements more consistent across RSL specifications. For example, for any RSL element, its ID, name, and description shall be consistent (i.e., must not include terms that are listed as alternatives in the glossary and use the main term instead). The presence of any inconsistent fragment is identified as a warning, whereas loops in relationships and duplicated IDs are identified as an error.

On the other hand, the definition of linguistic rules and association to RSL elements allows defining validations that make the IDs, names, and descriptions more consistent and complete across RSL specifications. For example, the actors' names shall follow the linguistic pattern "(Noun | ProperNoun)", and the descriptions of functional requirements shall follow the linguistic pattern ""System" + "shall"+ (Verb) + (DataEntity.name)".

#### 5.1.2 Document Automation

Upon validating the RSL specifications, we generated several documents based on these specifications. For example, to demonstrate the documentation automation process using

---
[10] https://github.com/peerigon/angular-expressions

templates discussed in section 4, we successfully generated the document illustrated in Figure 13.

This document the shows the stakeholder elements generated, each using the template fragment "Stakeholder {nameAlias} is a {type.type}" to indicate their name and type.

In the case of use case specifications, the template engine generates a table for each use case specifying its name, type, the reference to name of the primary actor and all associated actions.

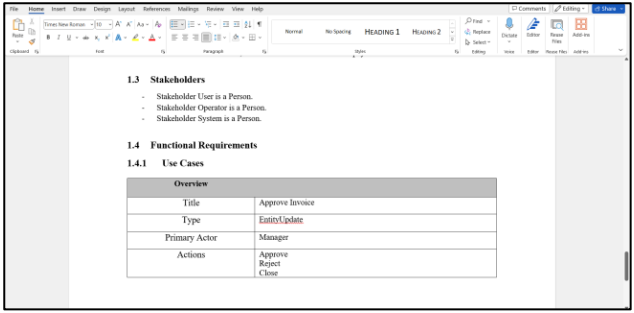

**Figure 13: Example of a document generated using a template file**

## 5.2 Usability Assessment

We conducted a usability assessment session in collaboration with Ricardo Silva [18], to assess the functionalities and usability of the ITOI platform and its support for the RSL language. Ricardo Silva's research introduces the ITOI platform, an online IDE integrated within ITLingo-Cloud to support the ITLingo languages like the RSL language. During this session, participants were asked to complete a set of tasks with the RSL language using the ITOI platform and then provide feedback by answering a questionnaire to assess the overall usability, and relevance of both ITOI and ITLingo RSL language. The questionnaire was answered by 16 users, 14 men and 2 women, with at least a BSc degree, namely 3 users with a BSc, 9 with an MSc, and 4 with a Ph.D.

Regarding the questions relevant to this research, users evaluated the easiness and usefulness of performing specific tasks, namely: (1) creating an RSL specification; (2) validating specifications based on grammar, custom checks, and linguistic rule checks; (3) importing elements from other packages and systems; and (4) generating files from RSL specifications. The results are summarized in Table 1, showing the average score for the easiness and usefulness of performing each task.

The user session evaluated the intuitiveness and usefulness of the proposed techniques, providing highly favorable results and valuable information to guide future implementations and improvements.

**Table 1: Overview of the easiness and usefulness of performing tasks**

| Task | Easiness (Average) | Usefulness (Average) |
|---|---|---|
| Creation of RSL specification | 4.06 | 4.19 |
| Error validation based on grammar checks | 3.94 | 3.94 |
| Error validation based on custom checks | 4.25 | 4.19 |
| Error validation based on linguistic rule checks | 3.88 | 4.31 |
| Import of elements from other packages and systems | 4.25 | 4.13 |
| Generation of files from RSL specifications | 4.38 | 4.56 |

## 6 DISCUSSION

This section discusses the research performed in the requirements validation based on NLP techniques and document automation process. It also presents some tools that support the validation of requirements using NLP methods and discusses the need for the type of advanced validations and document automation techniques as introduced in this paper.

### 6.1 Literature Review

Liu et al. [19] and Arora et al. [20] research aims to reduce common errors found in requirements specifications written in natural language by using NLP techniques to analyze requirements, checking for compliance with the defined templates, or detecting defects like ambiguities or incompleteness. However, our research demonstrates the advantages of using controlled natural languages for the analysis of requirements specifications and the identification of common errors.

In addition, we also promote validations based on NLP techniques with linguistic rules, giving more flexibility to those validations while, at the same time, becoming a learning tool for those authoring requirements, providing feedback to the user of any defects found.

Eito-Brun et al. [21] and Menezes et al. [22] have conducted research in the context of document automation. Both studies aim to reduce the time-consuming, and prone to error procedures involved in document creation when performed by a human. They explore the use of templates and structured languages to simplify the document production process, thereby minimizing errors in the final document.

Nonetheless, in our approach, we also introduce an additional layer of validation on the data that will be used to generate the document. This additional step guarantees that the final document is well-structured and based on information that has been validated, thus increasing its rigor and reliability. Moreover, our approach stands out from these studies due to its versatility in document automation techniques. This includes the generation of JSON files, to facilitate the transformation of requirements specifications into other data types, and the generation of documents in more commonly used formats (e.g., DOCX or PPTX) using template documents featuring flexible template tags based on RSL objects or tags based on some expressions with computational logic.

## 6.2 Related Tools

From another practical perspective, some companies have developed tools that use NLP techniques that analyze natural language requirements systematically and automatically, namely the following tools: RQA, RAT and QVscribe.

IBM Engineering Requirements Quality Assistant (RQA)[11] uses IBM Watson®[12] NLP tools pre-trained with INCOSE guidelines [12] to improve the quality of requirements by detecting potential ambiguities. The RQA assigns a score to each requirement to assess compliance with the guidelines.

The Requirements Authoring Tool (RAT)[13] helps engineers improve the text quality of documents and models, providing a standardized set of patterns that shall be used and that align with an organization's policies. RAT also generates quality reports that highlight requirements defects.

QVscribe[14] is a requirements analysis tool that can be integrated with other platforms (e.g., MS Word), and checks for compliance with the best requirements analysis practice standards such as INCOSE [12], and requirements writing best practices guidelines such as EARS [13]. It also analyses the requirement document and provides a quality score for each specific requirement sentence.

These tools support requirements authoring with NLP techniques that verify and impose linguistic rules defined by standards and guidelines while also serving as learning tools for requirements writers. Although they all support the configuration of which validations shall be performed, it is crucial to note that the RAT is the only one that allows the definition of such rules (that specifications must adhere to in order to achieve validity), by using a controlled natural language like RSL. However, it is worth mentioning that none of these tools support the validations based on customized checks and linguistic rules checks discussed in this research.

Table 2 compares the types of validation checks discussed in this research and related tools.

---

[11] https://www.ibm.com/docs/en/erqa?topic=engineering-requirements-quality-assistant
[12] https://www.ibm.com/watson
[13] https://www.reusecompany.com/rat-authoring-tools
[14] https://qracorp.com/requirements-software-tool-qvscribe-word-and-excel

**Table 2: Comparison of the validations discussed in this paper with similar tools**

| Tools | Grammar-based checks | Customized checks | Linguistic rules checks |
|---|---|---|---|
| RQA | No support | Identifies issues such as imprecise verbs, passive voice, and missing actors. | No support |
| RAT | Checks the structure and grammar of pre-defined requirements | Define pre-defined quality metrics that measure the level of inconsistency, ambiguity or duplicated items. | No support |
| QVscribe | No support | Allows to customize several problem types (e.g., passive voice, vague words) based on industry best practices. | No support |

## 7 CONCLUSION

This paper introduces new techniques for validating and generating documents based on RSL specifications. It presents validations based on customized and linguistic rule checks using the language's specific validator and the Compromise NLP library. In addition, it discusses document automation techniques based on valid RSL specifications, that generate documents like DOCX, PPTX, or alternative file formats, such as JSON or plain text.

## 7.1 Main Contributions

This research introduces new validation and document automation techniques to the RSL language, specifically to the ITLingo-Cloud environment. We accomplished integration with the ITLingo-Cloud system, specifically focusing on ITOI, as detailed in the research of Ricardo Silva [18], facilitating a web-based requirements management tool to promote collaboration and assist the authoring of requirements specifications.

The RSL language includes grammar validations as a standard check to guarantee that requirements specifications follow the language's syntax. Nevertheless, the validations based on customized checks and linguistic rules checks discussed in this paper allow identifying errors in the language that are not possible to identify with the grammatical validations, to further reduce possible inconsistency and ambiguity in RSL specifications.

In addition, the validation based on linguistic rules checks introduces a novel and flexible validation technique, allowing RSL users (e.g., requirements engineers) to define their own rules and guidelines for the IDs, names, and descriptions of requirements specifications that shall be followed, using elements from the language to define those rules.

Furthermore, the validation techniques discussed in this paper validate requirements and assist users in maintaining consistency and systematicity, while also offering feedback when a specification has a defect, suggesting quick fixes.

The document automation techniques discussed in this research are also essential in the process of defining requirements. Once the RSL specifications have been validated, they can be transformed into technical documents more commonly used by business stakeholders. These document automation techniques are flexible, particularly those using template files, facilitating the generation of technical documents with a predefined structure.

Therefore, both validation approaches discussed in this paper aim to reduce the time and effort required for manual validation and, as a result, generate technical documents from these validated RSL specifications.

## 7.2 Future Work

This paper presents an approach to improve the authoring of requirements specifications using validation and document automation techniques.

For future work, we intend to explore the combination of the techniques discussed in this research with "Generative AI" techniques, including: (a) extraction/conversion of natural language specifications into RSL to facilitate the validation process, as previously discussed in this research, and (b) generating RSL specifications in natural language to enhance their comprehension for users unfamiliar with RSL.

Finally, we aim to extend this research results: (1) with additional projects and research in either real or semi-real contexts, in order to validate the usefulness and practicality of the validation and document automation techniques; (2) into other ITLingo languages, such as ASL (Application Specification Language) [23], by integrating the document automation and validation techniques presented in this paper.